\documentclass{ws-p8-50x6-00}

\begin{document}
\def\pp{{\, \mid \hskip -1.5mm =}}
\def\cL{{\cal L}}
\def\be{\begin{equation}}
\def\ee{\end{equation}}
\def\bea{\begin{eqnarray}}
\def\eea{\end{eqnarray}}
\def\tr{{\rm tr}\, }
\def\nn{\nonumber \\}
\def\e{{\rm e}}
\def\D{{D \hskip -3mm /\,}}

\title{AdS Black Hole in $R^2$-Gravity}

\author{Sachiko Ogushi}

\address{Yukawa Institute for Theoretical Physics,\\ 
Kyoto University, Kyoto 606-8502, JAPAN\\ 
E-mail: ogushi@yukawa.kyoto-u.ac.jp}


\maketitle

\abstracts{
We discuss some properties of higher derivative (HD) bulk gravity 
without Riemann tensor square term.  
Such a kind of gravity admits Schwarzschild Anti de Sitter (SAdS) 
black hole as exact solution. 
It is shown that induced brane geometry on such background 
is Friedmann-Robertson-Walker (FRW) radiation dominated Universe. 
We show that HD terms contributions appear 
in the Hawking temperature, entropy and Hubble parameter via the 
redefinition of 5-dimensional gravitational constant and AdS scale parameter. 
These HD terms do not destroy the AdS-dual description of
radiation represented by strongly-coupled CFT. So-called 
Cardy-Verlinde formula which expresses cosmological 
entropy as square root from other parameters and 
entropies is also derived in $R^2$-gravity.  This talk is
based on works with Shin'ichi Nojiri and Sergei D. Odintsov\cite{sa}.}
\section{Introduction}
  
It has been realized recently\cite{sv} that brane equations 
of motion are exactly FRW equations with 
radiation matter which plays the role of CFT 
in AdS/CFT correspondence\cite{mm}.  
Furthermore, FRW equation can be rewritten in the form of
so-called Cardy-Verlinde formula\cite{ev} relating 
cosmological entropy with the one of CFT which exists on 
the boundary of AdS.  

We will be interested in the further study of the CFT 
dominated Universe as the brane in the background of 
HD gravity which is known to possess 
AdS Black Hole (BH) solution.
The interest in the HD bulk gravity 
is caused by the following: First of all, any effective 
stringy gravity includes HD terms of 
different order. Second, from the point of view of 
AdS/CFT correspondence the $R^2$-terms give next-to-leading 
terms in large $N$ expansion\cite{af} as it was directly 
checked in the calculation of holographic conformal anomaly
from bulk $R^2$-gravity\cite{sn}. 
Third, HD gravity
may serve as quite good candidate for the construction of
realistic brane-world cosmologies\cite{oo}.  
\section{AdS Black holes in bulk $R^2$-gravity, Surface terms}
Let us consider thermodynamics of AdS BH in bulk $R^2$-gravity.
The calculation of thermodynamical quantities like mass and entropy 
will be necessary to relate them with the corresponding ones in brane FRW 
Universe.  The general action of $d+1$-dimensional $R^2$-gravity is given by
\bea
\label{vi}
I=\int d^{d+1} x \sqrt{-\hat G}\left\{a \hat R^2 
+ b \hat R_{\mu\nu}\hat R^{\mu\nu}
+ \epsilon \hat R_{\mu\nu\xi\sigma}\hat R^{\mu\nu\xi\sigma}
+ {1 \over \kappa^2} \hat R - \Lambda \right\}\ .
\eea
When $\epsilon=0$ \footnote{For non-zero $\epsilon$ such S-AdS BH solution may be constructed perturbatively\cite{sa}.} 
Schwarzschild-anti de Sitter space is an exact solution:
\bea
\label{SAdS}
ds^2&=&\hat G_{\mu\nu}dx^\mu dx^\nu =
-\e^{2\rho}dt^2 + \e^{-2\rho}dr^2 
+ r^2\sum_{i,j}^{d-1} g_{ij}dx^i dx^j\ , \nn
\e^{2\rho}&=&{1 \over r^{d-2}}\left(-\mu + {kr^{d-2} \over d-2} 
+ {r^d \over l^2}\right)\, .
\eea
The curvatures have the form: $\hat R=-{d(d+1) \over l^2}\; , \; 
\hat R_{\mu\nu}= - {d \over l^2}\hat G_{\mu\nu}\ .$
In (\ref{SAdS}), $\mu$ is the parameter corresponding to mass 
and the scale parameter $l$ is given by solving the 
equation of motion of $I$ with respect to $\hat{G}_{\mu\nu}$:
\be
\label{ll}
0={d^2(d+1)(d-3) a \over l^4} + {d^2(d-3) b \over l^4} \nn
- {d(d-1) \over \kappa^2 l^2}-\Lambda\ .
\ee
We also assume $g_{ij}$ expresses the Einstein manifold, 
defined by $r_{ij}=kg_{ij}$, where $r_{ij}$ is the Ricci tensor 
given by $g_{ij}$ and $k$ is a constant\footnote{
For example, if $k>0$ the boundary can be 4-dimensional 
de Sitter space (sphere when Wick-rotated), if $k<0$, anti-de Sitter 
space or hyperboloid, or if $k=0$, flat space. 
By properly normalizing the coordinates, one can choose $k=2$, $0$, 
or $-2$.}.

The calculation of thermodynamical quantites like free 
energy $F$, the entropy ${\cal S}$ and the energy $E$ may be done 
with the help of the following method\cite{so}. After 
Wick-rotating the time variable by $t \to i\tau$, 
we assume $\tau$ has a period of ${1\over T}$.  
The free energy $F$ can be obtained from the action $I$ in (\ref{vi})
where the classical solution is substituted because
$F$ is related with $I$ as $F=-TI$.
Substituting Eqs.(\ref{ll}) into (\ref{vi}) in the case of $d=4$ 
with $\epsilon=0$, 
we obtain free energy $F$.  By using $F$,  
the entropy ${\cal S}$ and energy $E$ are given by
\bea
\label{ent}
{\cal S }&=& -{dF \over dT_H}
=-{dF \over dr_{H}}{dr_{H} \over dT_H}
={V_{3}\pi r_H^3 \over 2}
\left( {8 \over \kappa^2}- {320 a \over l^2}
 -{64 b \over l^2} \right) \, , \\
\label{ener}
E&=&F+T{\cal S} = {3V_{3}\mu \over 8}
\left( {8 \over \kappa^2}- {320 a \over l^2}
 -{64 b \over l^2} \right)\ .
\eea
Here $V_{3}$ is the volume of 3-dimensional sphere,  
$r_H$ is the horizon radius given by solving the 
equation $\e^{2\rho( r_H )}=0$, 
and $T_H$ is the Hawking temperature given by
$T_H ={1\over 4\pi}{d\over dr } \e^{2\rho} |_{r=r_{H}} \;$.  
The above equations reproduce the standard Einstein theory results when 
$a=b=0$.  Note that  one can consider the limit of 
$l\rightarrow 0$, where the background spacetime becomes 
 flat Minkowski space. Since the scalar curvature and Ricci 
tensor vanishes in the flat Minkowski, we cannot derive the 
thermodynamical quantities by evaluating the action $I$, 
which vanishes, if we start with the flat Minkowski background 
from the begining. Then finite $l$ 
would give a kind of the regularization. 

Before considering the dynamics of the brane, we review the 
problem of the variational principle in the Einstein gravity, 
whose action is given by
$ I_{\rm E}={1 \over \kappa^2} \int d^{d+1} x \sqrt{-\hat G}
\hat R \ .$
The scalar curvature contains the second order derivative 
of the metric tensor $\hat G_{\mu\nu}$ with respect the 
coordinates. Therefore if there is a boundary, which we denote 
by $B$, under the 
variation $\delta\hat G_{\mu\nu}$, $\delta S_{\rm E}$ 
contains, on the boundary, the derivative of 
$\delta\hat G_{\mu\nu}$ with respect to the coordinate 
perpendicular to the boundary, which makes the variational 
principle ill-defined. Therefore we need to add a surface 
term to the action, which is called  the Gibbons-Hawking 
surface term\cite{gh} :$I_{\rm GH}
={2 \over \kappa^2} \int_B d^d x \sqrt{-\hat g}\nabla_\mu n^\mu$,
where $n_\mu$ is the unit vector perpendicular to the boundary 
and $\hat g_{mn}$ is the boundary metric induced from $\hat G_{\mu\nu}$.  
For the $R^2$-gravity, it is presumably also possible 
to assumes the surface terms in the following forms:
\bea
\label{sf1}
I_{b}^{(1)}= {2\over \tilde{\kappa}^2} \int d^{d}x \sqrt{\hat{g}} 
\nabla_{\mu}n^{\mu}  \, , \quad
I_{b}^{(2)}= -\eta \int d^{d}x \sqrt{\hat{g}}\ .
\eea
The parameter $\eta$ (brane tension) which is usually 
free parameter in brane-world cosmology is not free any more 
and can be determined by the condition that 
the leading divergence of bulk AdS should vanish when one substitutes
the classical solution (\ref{SAdS}) 
into the action (\ref{vi}) with $\epsilon=0$ 
and into $I_{b}^{(1)}+I_{b}^{(2)}$.  
Then we obtains $\eta$. For later convenience, 
we choose the metric: 
\be
\label{S1}
ds^2=dq^2 -\e^{\zeta(q,\tau)}d\tau^2 + \e^{\xi(q,\tau)}
g_{ij}dx^i dx^j\ .
\ee  
The variation of the actions leads two kinds 
of equations of motion.  First equations of motion is 
derived from the condition that 
the coefficients in front of $\delta\zeta_{,q}$ and 
$\delta\xi_{,q}$ vanish. 
\be
\label{S4}
{1 \over \tilde\kappa^2}= {1 \over \kappa^2} 
 - {2d(d+1) a \over l^2 } - {2d b \over l^2}\ ,
\ee
And second equations of motion which is the original one
is obtained by the condition that the coefficients 
in front of $\delta\zeta$ and $\delta\xi$ vanish:
$ \left.\left(\zeta_{,q} + (d-1)\xi_{,q}\right)\right|_{q=0}
= \left.(d-1)\xi_{,q}\right|_{q=0}
= \tilde\kappa^2 \eta\ = {2(d-1) \over l}$.
We regard this equation is related to the dynamics of brane.  
Especially when $\e^\xi=\e^\zeta=l^2\e^{2A}$, we 
obtains the relation: 
$\left.A_{,q}\right|_{q=0}= {1 \over l}\ $.  
When $d=4$, by using Eq.(\ref{S4}),
the entropy (\ref{ent}) and the energy (\ref{ener}) are
rewritten by $\tilde{\kappa}$ as follows:
\bea
\label{ent2}
{\cal S }= {4V_{3}\pi r_H^3 \over \tilde\kappa^2}\ ,\quad
E={3V_{3}\mu \over \tilde\kappa^2}\ .
\eea
Therefore the corrections from the HD terms 
appear through the redefinition of  gravitational coupling $\kappa$ to 
$\tilde\kappa$ through (\ref{S4}) and $l$ given by (\ref{ll}).
\section{The FRW equations and Cardy-Verlinde formula}
Let us rewrite the metric (\ref{SAdS}) of SAdS 
in a form of (\ref{S1}) with $\e^\xi=\e^\zeta=l^2\e^{2A}$.  
It is possible to change the coordinates from $(t,r)$
to $(q,\tau)$ as the metric on the brane takes FRW form: 
$ds_{\rm brane}^2= -d \tilde t^2  + l^2\e^{2A}
\sum_{i,j=1}^{d-1}g_{ij}dx^i dx^j$, 
where we choose $r=l\e^A$ and $\tilde t$ which is
defined by $d\tilde t = l\e^A d\tau$.  
By using the relation between $(t,r)$
and $(q,\tau)$ to form FRW type metric, 
we have the square of the Hubble constant $H$ 
which is defined by $H={dA \over d\tilde t}$ and 
we can rewrite $H^2$ in the form of the 
FRW equation by using (\ref{SAdS}) and 
the relation $\left.A_{,q}\right|_{q=0}
= {1 \over l}\ $:
\bea
\label{e4}
H^2 = A_{,q}^2 - {\e^{2\rho}\e^{-2A} \over l^2}
= - {k \over (d-2)r^2} + {\kappa_d^2 \over (d-1)(d-2)}
{\tilde E \over V}\ ,
\eea
where ${\tilde E}={(d-1)(d-2) \mu V_{d-1} \over \kappa_d^2 r}$ and 
$V=r^{d-1}V_{d-1}\; $; $V_{d-1}$ is the volume of the 
$(d-1)$-dimensional sphere with a unit radius and 
$\kappa_d$ is the $d$-dimensional 
gravitational coupling, which is given by 
$\kappa_d^2={2 \tilde\kappa^2 \over l}\ .$  
By differentiating Eq.(\ref{e4}) with respect to $\tilde t$, 
since $H={1 \over r}{dr \over d\tilde t}$, we obtain the 
second FRW equation. From this equation, we can read 
the pressure of the matter on the brane as $p={(d-2)\mu \over r^d \kappa_d^2}\
$.  Thus we find the relation $0=-{\tilde E \over V} + (d-1)p\ ,$
which tells that the trace of the energy-stress tensor coming 
from the matter on the brane vanishes.  
Therefore the matter on the brane can be 
regarded as the radiation, i.e., the massless fields. 
In other words, field theory on the brane should be conformal one as in case 
of Einstein brane.  
When $d=4$, we find $\tilde E={l \over r}E\ .$
It should be noted that when $r$ is large, 
the metric (\ref{SAdS}) becomes the CFT metric 
which tells that the CFT time $\tilde t$ is equal 
to the AdS time $t$ times the factor ${r \over l}$, 
that is $t_{\rm CFT}={r \over l} t \;$.
Therefore the relation between $E$ and $\tilde{E}$ 
expresses that $\tilde{E}$ is the energy in CFT.  
Assuming AdS/CFT correspondence, the HD 
terms in (\ref{vi}) correspond to the $1/N$ corrections 
in the large $N$ limit of some gauge theory, which could 
be a CFT on the brane. For $\kappa^2 a$, 
$\kappa^2 b\ll 1$, we can rewrite $\tilde E$
using (\ref{S4}) as 
${\tilde E}\sim {2(d-1)(d-2) \kappa^2\mu V_{d-1} \over r}
\left( 1 + {2d(d+1)\kappa^2 a \over l^2 }
+ {2d b\kappa^2 \over l^2}\right)\ .$
Then the parameters $a$ and $b$ could express 
the $1/N$ correction of the next-to-leading order 
of $1/N$ expansion. 

Recently the FRW equation in $d$-dimension can be regarded 
as a $d$-dimensional analogue of
the Cardy formula of 2-dimensional CFT\cite{ev}:
\be
\label{CV1}
\tilde {\cal S}=2\pi \sqrt{
{c \over 6}\left(L_0 - {k \over d-2}{c \over 24}\right)}\ .
\ee
In the present case, 4-dimensional entropy ${\cal S}$
is obtained by identifying $
{2\pi \over 3}\tilde E r  \Rightarrow 2\pi L_0 \ ,\ 
{2V \over \kappa_4^2 r} \Rightarrow {c \over 24} \ ,\
{8\pi HV \over \kappa_4^2} \Rightarrow \tilde {\cal S}\ $.
Since the FRW-like equation (\ref{e4}) has the form (\ref{CV1}).  
Then one can evaluate holographic entropy 
$\tilde{\cal S}$ when the brane crosses the horizon $r=r_H$. When $r=r_H$, 
Eq.(\ref{e4}) tells that $H=\pm {1 \over l}$ 
where the plus sign corresponds to the expanding brane universe 
and the minus sign to the contracting one.  Taking 
the expanding case, we find 
$\tilde{\cal S}=4\pi r_H^{3} V_{3} /\tilde\kappa^2$. 
Thus we realize that the entropy $\tilde{\cal S}$ is 
identical with ${\cal S}$ in (\ref{ent}), which is nothing 
but the black hole entropy.  
We now stress again that compared with the Einstein gravity case, 
the corrections from the HD terms 
always appear through the redefinition of gravitational coupling 
$\kappa$ to $\tilde\kappa$ via (\ref{S4}) and when the 
length scale $l$ is given by (\ref{ll}). From the viewpoint of 
AdS/CFT correspondence, the HD terms 
correspond to the $1/N$ corrections in the large $N$ limit of 
some gauge theory, which could be a CFT on the brane.  
Then $\tilde{E}={l \over r} E$ and $\tilde{\cal S}={\cal S}$ 
would tell that AdS/CFT correspondence could be valid in 
the next-to-leading order of the $1/N$.  

\section*{Acknoweledgements} 
The author would like to thank S. Nojiri and S.D. Odintsov for
fruitful collaborations\cite{sa} which this talk based upon.
This work is supported in part by the Japan Society for the
Promotion of Science.

\end{document}